\documentclass[aps,prl,twocolumn,groupedaddress]{revtex4}

\usepackage{fancyhdr}
\usepackage{amssymb}
\usepackage{graphicx}
\usepackage[colorlinks,hyperfootnotes=false]{hyperref}
\usepackage{slashed}
\usepackage{ulem}

\usepackage{subfigure}
\usepackage{tikz}
\usetikzlibrary{positioning,arrows}
\usetikzlibrary{decorations.pathmorphing}
\usetikzlibrary{decorations.markings}

\newcommand{\eq}[1]{Eq.~(\ref{#1})}
\newcommand{\eqs}[2]{Eqs.~(\ref{#1}) and (\ref{#2})}
\newcommand{\eqss}[3]{Eqs.~(\ref{#1}), (\ref{#2}) and (\ref{#3})}

\newcommand{\GeV}{\mathinner{\mathrm{GeV}}}

\def\l{\left}
\def\r{\right}

\def\bea{\begin{eqnarray}}
\def\eea{\end{eqnarray}}
\def\beq{\begin{equation}}
\def\eeq{\end{equation}}

\begin{document}

%\preprint{}

\title{Local $Z_2$ scalar dark matter model confronting galactic ${\mathrm GeV}$-scale $\gamma$-ray}

%\title{Scalar dark matter stabilized by local $Z_2$ symmetry \\ and the INTEGRAL  511 keV $\gamma$ ray}


\author{Seungwon Baek}
\email[]{sbaek1560@gmail.com}
%\thanks{}
\affiliation{School of Physics, KIAS, Seoul 130-722, Korea}

\author{P. Ko}
\email[]{pko@kias.re.kr}
%\thanks{}
\affiliation{School of Physics, KIAS, Seoul 130-722, Korea}

\author{Wan-Il Park}
\email[]{wipark@kias.re.kr}
%\thanks{}
\affiliation{School of Physics, KIAS, Seoul 130-722, Korea}


\date{\today}

\begin{abstract}
We present a scalar dark matter (DM) model where DM ($X_I$) is stabilized by a local 
$Z_2$ symmetry originating from a spontaneously broken local dark $U(1)_X$. 
Compared with the usual scalar DM with a global $Z_2$ symmetry, the local $Z_2$ model possesses
three new extra fields, dark photon $Z^{'}$, dark Higgs $\phi$ and the excited partner of scalar DM ($X_R$), 
with the kinetic  mixing and Higgs portal interactions dictated by local dark gauge invariance. 
The resulting model can accommodate thermal relic density of scalar DM without 
conflict with the invisible Higgs branching ratio and the bounds from DM direct detections,  
thanks to the newly opened channels, $X_I X_I \rightarrow Z^{'} Z^{'}, \phi\phi$.
In particular, due to the new particles, the ${\rm GeV}$ scale $\gamma$-ray 
excess from the Galactic Center (GC) can be originated from the decay of dark Higgs boson which is produced in DM annihilations. 
%Also the  muon $(g-2)$ anomaly can be explained if the mass of dark photon is around $\sim 20~ {\rm MeV}$ with the kinetic mixing of $\mathcal{O}(10^{-3})$.   
%Moreover, with a $U(1)_X$-charged $\keV$ scale sterile neutrino introduced in addition to %right-handed neutrinos, the model can also accommodate not only the muon $(g-2)$ for light dark %photon $m_{Z^{'}} \sim 20-60$ MeV and kinetic mixing $\epsilon \sim 10^{-3}$ but also $3.55 \keV$ %$X$-ray line hinted by recent observations.
\end{abstract}

\pacs{}
%\keywords{}

\maketitle

\section{Introduction}

One of the great mysteries of particle physics and cosmology is the so-called nonbaryonic  
dark matter (DM) which occupies about $27$ \% of the energy density of the present universe
\cite{Planck1,Planck2}.   DM particle should be very long-lived or absolutely stable, and interact
with photon or gluon very weakly ( i.e. at least no renormalizable interaction), but otherwise its  
properties are largely unknown. 

The simplest DM model is the real scalar DM model described by the % following 
Lagrangian~\cite{,Silveira:1985rk,Burgess:2000yq,Cline:2013gha,McDonald:1993ex}: 
\begin{equation}
{\cal L}_{\rm DM} = \frac{1}{2} \partial_\mu S \partial^\mu S - \frac{m_S^2}{2}  S^2 - 
\frac{\lambda_{HS}}{2} S^2 H^\dagger H - \frac{\lambda_S}{4 !} S^4 ,
\end{equation}
with $Z_2$ symmetry ($S \rightarrow -S$). 
This model has been studied extensively in literature, and could be considered as a canonical
model for non-supersymmetric DM. 
However $Z_2$ symmetry in Eq.~(1) is not usually specified whether it is global or local. 
If it were global, it may be broken by gravity effects, described by higher dimensional 
nonrenormalizable operators such as 
\[
{\cal L}_{{Z_2}\rm breaking} = \frac{c_5}{M_{\rm Planck}} S O_{\rm SM}^{(4)}
\]
where $O_{\rm SM}^{(4)}$ is any dim-4 operator in the Standard Model (SM) such as 
$G_{\mu\nu}G^{\mu\nu}$ or Yukawa interactions, etc.  Such dim-5 operators will make the scalar DM
$S$ decay immediately unless its mass is very light $\lesssim O(1)$ keV if we assume 
$c_5 \sim O(1)$~\cite{Baek:2013qwa}.   Therefore global $Z_2$ would not be enough to stabilize or 
make the weak scale DM $S$ long-lived enough.  Therefore it would be better to use local $Z_2$ symmetry 
to stabilize weak scale DM~\cite{Baek:2013qwa}. 

This new local gauge symmetry has another nice feature that DM also has its own gauge 
interaction just as all the SM particles do  feel some gauge interactions, with a possibility
of strong self interaction for light dark gauge bosons and/or dark Higgs~\cite{Ko:2014nha}.
Dark gauge symmetry can be realized naturally in  
superstring theory, for example, where the original gauge group with a huge rank is broken 
into $G_{\rm SM} \times G_{\rm Dark}$.

In this letter, we propose a simple scalar dark matter model based on a local $Z_2$ discrete symmetry 
originating from a spontaneously broken local $U(1)_X$, and investigate its phenomenology including relic density, 
possibilities of direct/indirect detections and addressing $\GeV$ scale $\gamma$-ray excess in Fermi-LAT $\gamma$-ray data in the direction of the Galactic Center (GC).  
In our local $Z_2$ model, there are 3 new extra fields (dark Higgs $\phi$,  dark photon $Z^{'}$,  
and unstable excited dark scalar $X_R$) dictated by local $U(1)_X$ dark gauge symmetry.
Due to the additional fields and presumed local dark gauge symmetry, the phenomenology of dark matter 
is expected to be distinctly different from the usual $Z_2$ scalar DM model described by Eq.~(1).
   
\section{Model}

Let us assume the dark sector has a local $U(1)_X$ gauge symmetry with scalar dark matter 
$X$ and dark Higgs $\phi$ with $U(1)_X$ charges equal to $q_X(X,\phi) = (1,2)$
~\cite{Krauss:1988zc}. The local 
$U(1)_X$ is spontaneously broken into a local $Z_2$ subgroup by nonzero VEV of $\phi$, $v_\phi$.  
Then the model  Lagrangian which is invariant under local dark gauge symmetry is given by 
\begin{widetext}
\begin{eqnarray}
{\cal L} & = & {\cal L}_{\rm SM}  
- \frac{1}{4} \hat{X}_{\mu\nu} \hat{X}^{\mu\nu} - \frac{1}{2} \sin \epsilon \hat{X}_{\mu\nu} \hat{B}^{\mu\nu} + D_\mu \phi^\dagger D^\mu \phi + D_\mu X^\dagger D^\mu X %+ \bar{\psi} \l( i \slashed{D} - m_\psi \r) \psi + \overline{N_R^c} \l( i\slashed{\partial} - M \r) N_R
 - m_X^2 X^\dagger X + m_\phi^2 \phi^\dag \phi
 \nonumber \\
 &&  
- \lambda_\phi \left( \phi^\dagger \phi \right)^2 - \lambda_X \left( X^\dagger X \right)^2\
- \lambda_{\phi X} X^\dagger X \phi^\dagger \phi
- \lambda_{\phi H} \phi^\dagger \phi H^\dagger H
- \lambda_{HX} X^\dagger X H^\dagger H  -  \mu \left( X^2 \phi^\dagger +  H.c. \right) . 
\label{eq:model}
\end{eqnarray}
\end{widetext}
We assume all $\lambda$'s and $\mu$ are positive, and the covariant derivative associated with the gauge field 
$\hat{X}^\mu$ is defined as $D_\mu \equiv \partial_\mu - iq_Xg_X \hat{X}_\mu$ with $g_X$ being the strength 
of $U(1)_X$ gauge interaction.  We have kept renormalizable operators only, assuming the effects from 
nonrenormalizable operators are negligibly small.  

Once the $U(1)_X$ symmetry is broken by nonzero VEV of $\phi$, we can replace 
$ \phi  \rightarrow ( v_\phi + \phi ) / \sqrt{2}$. Then the $\mu-$term becomes 
\[
\mu \l( X^2 \phi^\dag + H.c. \r)  = \frac{1}{\sqrt{2}} \mu v_\phi \l( X_R^2 - X_I^2 \r) ( 1 + 
\frac{\phi}{v_\phi}) ,
\]
with $X = \l(X_R + i X_I \r)/\sqrt{2}$, and  generates the mass splitting between $X_R$ and $X_I$, breaking $U(1)_X$ into $Z_2$  
under which  $X_{I,R}$ are odd and all the other fields are even. 
Note that the local $Z_2$ symmetry guarantees the stability of the dark matter
even if we consider Planck-scale-suppressed nonrenormalizable operators. 

The local $Z_2$ symmetry requires extra new fields (dark Higgs $\phi$ and  dark photon $Z_\mu^{'}$ 
(that mainly comes from $\hat{X}^\mu$),  as well as an excited partner of DM, $X_R$), compared with 
a singlet scalar dark matter model with an unbroken global $Z_2$ symmetry described by Eq. (1).  
These three new fields play important roles in DM phenomenology, phenomenological results  of which are 
qualitatively different from those in  the usual $Z_2$ scalar DM model. 
In particular, if we replace the dark Higgs field $\phi$ by its VEV and ignore the 
dark Higgs degree of freedom, our model becomes exactly the same as the excited scalar DM  model  
which was discussed in the context of 511 keV gamma ray and PAMELA positron excess
~\cite{Cholis:2008qq,Pospelov:2007xh}. 
The main difference of our model from the usual excited scalar  DM model is the presence of dark
Higgs field, which is dynamical and would change DM phenomenology completely.  
For example, the annihilation of DM for a right amount of thermal relic density can be dominated  by $X_I X_I \rightarrow \phi\phi$ and not by 
$X_I X_I \rightarrow Z' Z'$,  unlike the usual excited DM models.  
Details of this and related issues will be discussed elsewhere.

The $U(1)$ gauge kinetic mixing term can be diagonalized by the following transformation \cite{Babu:1997st}: 
\beq
\l(
\begin{array}{c}
\hat{B}_\mu \\ \hat{X}_\mu
\end{array}
\r)
=
\l(
\begin{array}{cc}
1 & - \tan \epsilon
\\
0 & 1/\cos \epsilon
\end{array}
\r)
\l(
\begin{array}{c}
B_\mu \\ \tilde{X}_\mu
\end{array}
\r)
\eeq
Diagonalizing the mass matrix subsequently, one then finds
\bea
\hat{B}_\mu &=& c_W A_\mu - \l( t_\epsilon s_\zeta + s_W c_\zeta \r) Z_\mu + \l( s_W s_\zeta - t_\epsilon c_\zeta \r) Z'_\mu \ ,
\nonumber \\
\hat{X}_\mu &=& \frac{s_\zeta}{c_\epsilon} Z_\mu + \frac{c_\zeta}{c_\epsilon} Z'_\mu \ ,
\\ 
\hat{W}_\mu &=& s_W A_\mu + c_W c_\zeta Z_\mu - c_W s_\zeta Z'_\mu \ .
\nonumber 
\eea
Here $s_W(c_W) = \sin \theta_W (\cos \theta_W)$ with $\theta_W$ being the weak mixing angle, and $\zeta$ is defined as
\beq
\tan 2 \zeta \equiv - \frac{m_{\hat{Z}}^2 s_W \sin 2 \epsilon}{m_{\hat{X}}^2 - m_{\hat{Z}}^2 \l( c_\epsilon^2 - s_\epsilon^2 s_W^2 \r)}
\eeq
where $m_{\hat{Z}}^2$ and $m_{\hat{X}}^2$ are the mass-squared of SM 
$Z$-boson and $\hat{X}_\mu$ respectively, before diagonalization of kinetic and mass terms.
In the limit of small kinetic mixing ($\epsilon \ll 1$) and $m_{\hat{X}}^2 \ll m_{\hat{Z}}^2$ which we are 
interested in, we find $t_\zeta \simeq s_W t_\epsilon$.  A summary of various constraints on $Z_\mu'$ 
can be found in Refs.~\cite{Chun:2010ve,Essig:2013lka}.

From the model Lagrangian Eq.~(\ref{eq:model}), we can work out the particle 
spectra at the tree level:
\begin{eqnarray}
m_{Z'}^2 & = & 4 g_X^2 v_\phi^2 ,   \nonumber \\
m_{R}^2 & = & m_X^2 + \sqrt{2} \mu v_\phi + \frac{1}{2} \lambda_{HX} v_H^2 + \frac{1}{2} \lambda_{\phi X} v_\phi^2 \\
m_{I}^2 & = & m_X^2 - \sqrt{2} \mu v_\phi + \frac{1}{2} \lambda_{HX} v_H^2 + \frac{1}{2} \lambda_{\phi X} v_\phi^2  \nonumber 
\end{eqnarray}
which show that the dark matter in our scenario is $X_I$.
In the true vacuum, 
%the vanishing tadpole conditions should be satisfied:  
%\bea
%m_\phi^2 &=& \frac{1}{2} \lambda_{\phi H} v_H^2 + \lambda_\phi v_\phi^2 ,
%\\
%m_H^2 &=& \frac{1}{2} \lambda_{\phi H} v_\phi^2 + \lambda_H v_H^2
%\eea 
%%where $\langle \phi \rangle = v_\phi /\sqrt{2}$ is the VEV of $\phi$.
%Because of non-zero $v_\phi$, there is a mixing between SM Higgs and $\phi$.
the mass matrix elements of Higgs fields are 
\bea \label{mass-m}
m_{hh}^2 &=& 2 \lambda_H v_H^2
 \nonumber   \\
m_{\phi h}^2 &=& \lambda_{\phi H} v_\phi v_H
\\
m_{\phi \phi}^2 &=& 2 \lambda_\phi v_\phi^2
 \nonumber 
\eea
where $v_H=246 \GeV$ is the VEV of SM Higgs.
The mass eigenvalues are
\beq
m_{1,2}^2 = \frac{1}{2} \l[ \l( m_{hh}^2 + m_{\phi \phi}^2 \r) \mp \sqrt{\l( m_{hh}^2 - m_{\phi \phi}^2 \r)^2 + 4 m_{\phi h}^4 } \r]
\eeq
Requiring $m_{1,2}^2 > 0$, one finds
\beq \label{stability-cond}
\l| \lambda_{\phi H} \r| < 2 \sqrt{\lambda_H \lambda_\phi}
\eeq
Interaction eigenstates can be expressed in terms of mass eigenstates as
\begin{equation}
\left( \begin{array}{c} h \\ \phi \end{array} \right)
=
\left( \begin{array}{cc} \cos \alpha & -\sin \alpha \\ \sin \alpha & \cos \alpha \end{array} \right)
\left( \begin{array}{c} H_2 \\ H_1 \end{array} \right)
\end{equation}
where the mixing angle $\alpha$ is defined as
\beq
\tan 2 \alpha = \frac{2 m_{\phi h}^2}{m_{hh}^2 - m_{\phi \phi}^2} .
\eeq
In the small mixing angle limit which we will be considering in this work, we have 
$H_2 \simeq h$ and $H_1 \simeq \phi$. 

\begin{table*}[t]
\begin{center}
\begin{tabular}{|c||c|c|c|c|c|c|c|c|c|c|c|}
\hline
Parameters & $\epsilon$ & $m_{Z'}$ & $m_I$ & $\mu$ & $m_1$ & $m_2$ & $v_\phi$ & $v_H$ & $\alpha$ & $\lambda_{\phi X}$ & $\lambda_{HX}$ \\
\hline \hline
Ranges & $\lesssim 10^{-3}$ & $\mathcal{O}(1)$ & $\mathcal{O}(10-100)$ & $\mathcal{O}(10)$ & $\sim m_I$ & $125$ & $\mathcal{O}(100)$ & $246$ & $\mathcal{O}(0.1)$ & $\mathcal{O}(10^{-3}-1)$ & $\mathcal{O}(10^{-3}-1)$ \\
\hline 
\end{tabular}
\end{center}
\caption{Free parameters and their ranges of consideration.
Dimensionful parameters are in $\GeV$ unit.
The value of $\lambda_X$ is not specified except the requirement of positivity.}
\label{tab:para}
\end{table*}
There are 12 free parameters in the local $Z_2$ scalar DM model as a whole:
\bea
&\epsilon, g_X,& 
\nonumber \\
&m_X, m_\phi, m_H, \mu,& 
\nonumber \\
&\lambda_X, \lambda_\phi, \lambda_H, \lambda_{\phi X}, \lambda_{HX}, \lambda_{\phi H}&
\eea
Among these, parameters associated with the Higgs sector are related as follows:
\beq
m_\phi, m_H, \lambda_\phi, \lambda_H, \lambda_{\phi H} \rightarrow v_\phi, v_H, \alpha, m_1, m_2
\eeq
For given $v_\phi$ and $v_H$, dark scalar masses are related as
\bea
m_X^2, \lambda_{\phi X}, \lambda_{H X} &\rightarrow& m_R^2 + m_I^2
\\
\mu &\rightarrow& m_R^2-m_I^2
\eea
Hence, for fixed $m_I$ and $m_R$ we can freely adjust $\lambda_{\phi X}$ and $\lambda_{HX}$ which 
will affect dark matter phenomenology.
The choice of values (or ranges) of these parameters is shown in Table~\ref{tab:para}, and the reason will become clear in the following sections.

\section{Dark Matter Phenomenology}

From now on, we denote $m_{1,2}$ as $m_{\phi, h}$ and assume 
\bea \label{mI-mphi}
30 \GeV \lesssim m_I \sim \ m_\phi \lesssim 80 \GeV
\eea
as the relevant range for the $\GeV$ scale $\gamma$-ray excess in the direction of GC. 
We also assume $g_X$ is somewhat small, for example, $\mathcal{O}(10^{-2})$ to provide a simple and 
clear picture of our scenario.  
The mass range of \eq{mI-mphi} implies $v_\phi \gtrsim \mathcal{O}(100) \GeV$ for $\lambda_\phi \lesssim 1$ from \eq{mass-m}, and $m_Z' = \mathcal{O}(1) \GeV$.

Our model allows tree-level dark matter self-interactions mediated by dark photon and 
scalar particles $H_{1,2}$ coming from $\phi$ and SM Higgs, for suitable choice of their 
masses and couplings  (see Ref.~\cite{Ko:2014nha} for  DM self-interactions in the scalar 
DM model with  local $Z_3$ symmetry).
However, for $m_I, m_\phi$ and $m_{Z'}$ in the ranges of our interest, the effects of DM self-interactions are negligible and  do not impose any meaningful constraint on $\alpha_X$, and we can ignore them.   

\subsection{Relic density}
If kinematically allowed, DM can annihilate to dark photon, non-SM Higgs and SM particles.
The Feynman diagrams for $X_I X_I \rightarrow Z^{'} Z^{'}$ are shown in Fig~\ref{fig:dm-to-2gammaP}.
%------------------
\tikzset{
particle/.style={thick,draw=black, postaction={decorate},
    decoration={markings,mark=at position .5 with {\arrow{triangle 45}}}},
fermion/.style={thick,draw=black, postaction={decorate},
    decoration={markings,mark=at position .5 with {\arrow{triangle 45}}}},
scalar/.style={thick,dashed,draw=black},
photon/.style={decorate, draw=black,
    decoration={coil,aspect=0}}
}
\begin{figure}
\centering
\subfigure[]{
\begin{tikzpicture}[node distance=0.8cm and 0.8cm]
%\coordinate[label=left:$X_I$] (e1);
%\coordinate[below right=of e1] (aux1);
%\coordinate[below left=of aux1,label=left:$X_I$] (e3);
%\coordinate[above right=of aux1,label=right:$X^{\mu}$] (e2);
%\coordinate[below right=of aux1,label=right:$X^{\nu}$] (e4);
\coordinate[] (e1);
\coordinate[below right=of e1] (aux1);
\coordinate[below left=of aux1] (e3);
\coordinate[above right=of aux1] (e2);
\coordinate[below right=of aux1] (e4);

\draw[scalar] (e1) -- (aux1);
\draw[scalar] (e3) -- (aux1);
\draw[photon] (aux1) -- (e2);
\draw[photon] (aux1) -- (e4);
\end{tikzpicture}
}
%\hspace{3mm}
%
\subfigure[]{
\begin{tikzpicture}[node distance=0.4cm and 0.8cm]
%\coordinate[label=left:$X_I$] (e1);
%\coordinate[below right=of e1] (aux1);
%\coordinate[above right=of aux1,label=right:$X^{\mu}$] (e2);
%\coordinate[below=0.8cm of aux1] (aux2);
%\coordinate[below left=of aux2,label=left:$X_I$] (e3);
%\coordinate[below right=of aux2,label=right:$X^{\nu}$] (e4);
\coordinate[] (e1);
\coordinate[below right=of e1] (aux1);
\coordinate[above right=of aux1] (e2);
\coordinate[below=0.8cm of aux1] (aux2);
\coordinate[below left=of aux2] (e3);
\coordinate[below right=of aux2] (e4);

\draw[scalar] (e1) -- (aux1);
\draw[photon] (aux1) -- (e2);
\draw[scalar] (e3) -- (aux2);
\draw[photon] (aux2) -- (e4);
\draw[scalar] (aux1) --  node[] {} (aux2);
\end{tikzpicture}
}
%\hspace{3mm}
%
\subfigure[]{
\begin{tikzpicture}[node distance=0.4cm and 0.8cm]
%\coordinate[label=left:$X_I$] (e1);
%\coordinate[below right=of e1] (aux1);
%\coordinate[above right=of aux1,label=right:$X^{\mu}$] (e2);
%\coordinate[below=0.8cm of aux1] (aux2);
%\coordinate[below left=of aux2,label=left:$X_I$] (e3);
%\coordinate[below right=of aux2,label=right:$X^{\nu}$] (e4);
\coordinate[] (e1);
\coordinate[below right=of e1] (aux1);
\coordinate[above right=of aux1] (e2);
\coordinate[below=0.8cm of aux1] (aux2);
\coordinate[below left=of aux2] (e3);
\coordinate[below right=of aux2] (e4);

\draw[scalar] (e1) -- (aux1);
\draw[photon] (aux1) -- (e4);
\draw[scalar] (e3) -- (aux2);
\draw[photon] (aux2) -- (e2);
\draw[scalar] (aux1) --  node[] {} (aux2);
\end{tikzpicture}
}
%\hspace{3mm}
%
\subfigure[]{
\begin{tikzpicture}[node distance=0.8cm and 0.4cm]
%\coordinate[label=left:$X_I$] (e1);
%\coordinate[below right=of e1] (aux1);
%\coordinate[below left=of aux1,label=left:$X_I$] (e3);
%\coordinate[right=0.8cm of aux1] (aux2);
%\coordinate[above right=of aux2,label=right:$X^{\mu}$] (e2);
%\coordinate[below right=of aux2,label=right:$X^{\nu}$] (e4);
\coordinate[] (e1);
\coordinate[below right=of e1] (aux1);
\coordinate[below left=of aux1] (e3);
\coordinate[right=0.8cm of aux1] (aux2);
\coordinate[above right=of aux2] (e2);
\coordinate[below right=of aux2] (e4);

\draw[scalar] (e1) -- (aux1);
\draw[scalar] (e3) -- (aux1);
\draw[photon] (aux2) -- (e2);
\draw[photon] (aux2) -- (e4);
\draw[scalar] (aux1) -- node[label=above:$H_{1,2}$] {} (aux2);
\end{tikzpicture}
}
\caption{\label{fig:dm-to-2gammaP} DM annihilations to two dark photons.}
\end{figure}
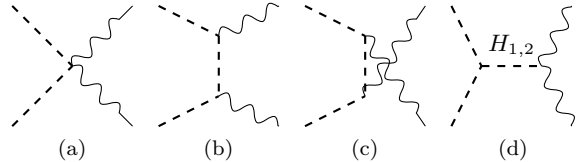
%---------------------------
When $g_X$ is small, the first three diagrams in Fig.~1 give thermal cross section which is too small 
to saturate canonical thermal cross section ($\langle \sigma v_{\rm rel} \rangle_{\rm th} \equiv 
3 \times 10^{-26} {\rm cm}^3/{\rm s}$). 
% (namely, light $Z^{'}$ limit). 
%{\color{red} annihilation  amplitude ($A$) $\rightarrow$ cross section} 
%with polarizations summed up as
%\beq \label{A-abc}
%A = \frac{1}{4} \frac{m_{Z'}^4 \l[ \l( s - 4 m_I^2 \r)^2 + 16 m_I^4 - 16 m_I^2 m_{Z'}^2 + 6 m_{Z'}^4 \r]}{\l( s %- 2 m_{Z'}^2 \r) v_\phi^4 }
%\eeq
%Note that $A$ goes to zero in the limit of $g_X \to 0$.
However, in the presence of the $s$-channel diagram (d), the scattering amplitude is finite even if 
$g_X=0$ because of the longitudinal component of $Z'$, and 
%Including diagram (d), the full form of the scattering amplitude becomes very complicated.  
%However, as can be seen from \eq{A-abc}, we observe that for particle spectra of our interest the contributions of %(a), (b) and (c) to the thermally averaged annihilation cross section are too small to saturate the canonical value. 
%Hence, 
only the diagram (d) becomes relevant.
In this case, ignoring the mass of dark photon in the  final states, one finds that the DM annihilation cross section
is approximately given by 
\beq \label{sv-darkphoton}
\langle \sigma v_{\rm rel} \rangle_{Z' Z'} \approx \frac{1}{8 \pi} \frac{s}{v_\phi^2} \l| \frac{\lambda_1 c_\alpha}{s - m_\phi^2 + i \Gamma_\phi m_\phi} + \frac{\lambda_2 s_\alpha}{s-m_h^2 + i \Gamma_h m_h} \r|^2
\eeq
where 
\bea \label{lambda1}
\lambda_1 &=& \l( \lambda_{\phi X} v_\phi - \sqrt{2} \mu \r) c_\alpha - \lambda_{HX} v_H s_\alpha
\\ \label{lambda2}
\lambda_2 &=& \l( \lambda_{\phi X} v_\phi - \sqrt{2} \mu \r) s_\alpha + \lambda_{HX} v_H c_\alpha
\eea
and $\Gamma_\phi$ and $\Gamma_h$ are the decay rates of $H_{1,2}$, respectively.

In case of two $H_1 (\simeq \phi)$ productions ($X_I X_I \rightarrow \phi\phi$), 
we take the small mixing angle limit again. 
%the amplitude of annihilation is 
%\beq
%\l| \mathcal{M} \r|^2 = \l| \sum_i \frac{\lambda_i \lambda_{i11}}{s - m_i^2 + i \Gamma_i m_i} + %\frac{\lambda_1^2}{t - m_I^2} + \frac{\lambda_1^2}{u-m_I^2} + \bar{\lambda} \r|^2
%\eeq
%where
%\bea
%\lambda_{111} &=& 3 v_\phi c_\alpha \l( 2 \lambda_\phi c_\alpha^2 + \lambda_{\phi H} s_\alpha^2 \r) 
%\nonumber \\
%&-& 3 v_H s_\alpha \l( 2 \lambda_H s_\alpha^2 + \lambda_{\phi H} c_\alpha^2 \r)
%\\
%\lambda_{211} &=& v_\phi s_\alpha \l[ 2 \l( 3 \lambda_\phi - \lambda_{\phi H} \r) c_\alpha^2 + \lambda_{\phi %H} s_\alpha^2 \r]
%\nonumber \\
%&+& v_H c_\alpha \l[ 2 \l( 3 \lambda_H - \lambda_{\phi H} \r) s_\alpha^2 + \lambda_{\phi H} c_\alpha^2 \r]
%\\
%\bar{\lambda} &=& \lambda_{\phi X} c_\alpha^2 + \lambda_{HX} s_\alpha^2
%\eea
%Note that in the small mixing limit,
%\beq
%\alpha \approx \frac{\lambda_{\phi H} v_\phi v_H}{2 \lambda_H v_H^2 - 2 \lambda_\phi v_\phi^2} \ll 1
%\eeq
%Hence, for $v_\phi \sim v_H$ and $\lambda_\phi \sim \lambda_H$, one finds 
%\beq
%\lambda_{\phi H} \ll \lambda_H, \lambda_\phi
%\eeq
%and 
%\beq
%\lambda_{111} \sim 6 \lambda_\phi v_\phi c_\alpha^3 \gg \lambda_{211} \sim \lambda_{\phi H} v_H c_\alpha^3
%\eeq
For a reasonable choice of parameters (e.g., $\lambda_{\phi H} \ll \lambda_H \sim \lambda_\phi \sim 0.1$ and $m_\phi < m_I$), as long as $m_I$ is far away from the $s$-channel resonance band, one finds that the contact interaction dominates DM annihilation into $\phi\phi$, and we get 
%In this case DM annihilation cross section is approximated to %($H_1 \simeq \phi$)
\bea \label{sv-Higgs}
\langle \sigma v_{\rm rel} \rangle_{\phi \phi}
&\simeq& \frac{1}{64 \pi m_I^2} \l( \lambda_{\phi X} c_\alpha^2 + \lambda_{HX} s_\alpha^2 \r)^2 \beta_\phi
 \\
&\simeq& \frac{2.46 \times 10^{-9}}{\GeV^2} \l( \frac{\lambda_{\phi X}}{0.07} \r)^2 \l( \frac{100 \GeV}{m_I} \r)^2 \beta_\phi \ .
\nonumber
\eea
Here $\beta_\phi \equiv \sqrt{1 - 4 m_\phi^2 /s}$ and we have used $\lambda_{H X} = 0.1$ 
and $\alpha=0.1$ in the second line.

DM can also annihilate directly to SM particles.
For $m_I$ in the range of our interest, the thermally-averaged annihilation cross section is 
\bea
\langle \sigma v_{\rm rel} \rangle_{f\bar{f}} 
&\simeq& \sum_f \frac{N_{c,f}}{4 \pi} \l( \frac{s}{4 m_I^2} \r)^{1/2} 
\nonumber \\
&\times& \l| \sum_i \frac{\lambda_i \lambda_{if}}{s - m_i^2 + i m_i \Gamma_i} \r|^2 \l( 1 - \frac{4 m_f^2}{s} \r)^{3/2}
\eea
where $N_{c,f}$ is the color factor, $\lambda_{1f} = -\sqrt{2} (m_f/v_H) s_\alpha$ and $\lambda_{2f} = \sqrt{2} (m_f/v_H) c_\alpha$ with $m_f$ being the mass of SM fermion $f$. 

%----------------------
\begin{figure}[ht]
\includegraphics[width=0.40\textwidth]{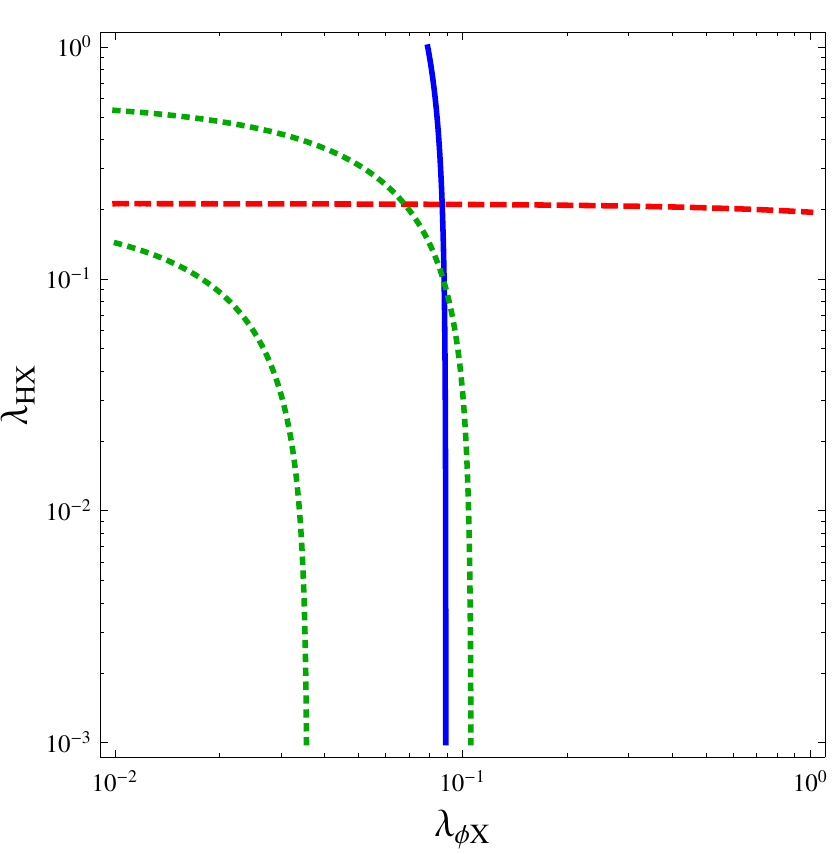}
\caption{\label{fig:svgg}
%\textit{Upper}: 
Contours satisfying $\langle \sigma v_{\rm rel} \rangle_i = \langle \sigma v_{\rm rel} \rangle_{\rm th}$ ($i=Z'Z', f\bar{f}, \phi\phi$) 
as functions of $\lambda_{\phi X}$ and $\lambda_{HX}$ for $\alpha=0.1, \ m_I=80 \GeV , 
\ m_\phi = 75 \GeV$, $v_\phi = 100 \GeV$, and $\mu=5 \GeV$.
Dotted green, dashed red, and solid blue lines are for $X_I X_I \to Z'Z', f \bar{f}, \phi \phi$, respectively.
$\langle \sigma v_{\rm rel} \rangle_i < \langle \sigma v_{\rm rel} \rangle_{\rm th}$ in the region between green lines, below red line, and left of the blue line, respectively.
%
%We used $\mu=1 \GeV$ for SM channel whose $\mu$-dependence is very weak and can be ignored.
% %\textit{Lower}: 
%Total DM annihilation cross section ($\langle \sigma v_{\rm rel} \rangle_{\rm tot} \equiv \langle \sigma v_{\rm rel} %\rangle_{Z'Z'} + \langle \sigma v_{\rm rel} \rangle_{\phi\phi} + \langle \sigma v_{\rm rel} \rangle_{f\bar{f}}$) 
%for for $\alpha=0.1, \ m_\phi = 75 \GeV, \  m_I=80 \GeV$, and $v_\phi = 100 \GeV$.  
%Solid red, dashed blue, dot-dashed green and dotted black lines correspond to $\mu=1,3,5,7 \GeV$, respectively. 
}
\end{figure}
%%-----------------------
In Fig.~2, the contour(s) of $\langle \sigma v_{\rm rel} \rangle = \langle \sigma v_{\rm rel} \rangle_{\rm th}$ for each of annihilation channel is shown in the plane of ($\lambda_{\phi X}, \lambda_{HX}$). % as the regions between solid ($\mu=1 \GeV$) lines or dashed ($\mu=10 \GeV$) lines.
As shown in the figure, the annihilation cross sections of all three channels ($X_I X_I \to Z'Z'/\phi\phi/f\bar{f}$) 
can be comparable to $\langle \sigma v_{\rm rel} \rangle_{\rm th}$ if either $\lambda_{\phi X}$ or $\lambda_{H X}$ is of $\mathcal{O}(0.1)$.
Interestingly, for $\langle \sigma v_{\rm rel} \rangle_{Z'Z'}$ with $\mu \sim 5 \GeV$, a cancellation between the contribution of $\lambda_{\phi X}$ and $\mu$ in \eqs{lambda1}{lambda2} results in an appearance of a band of $\langle \sigma v_{\rm rel} \rangle_{Z'Z'} < \langle \sigma v_{\rm rel} \rangle_{\rm th}$.
For a much smaller or larger $\mu$, such a band disappears for $\lambda_{\phi X}, \lambda_{HX} \lesssim 1$. 

If $X_R$ and $X_I$ are highly degenerate, the co-annihilation of $X_R$ and $X_I$ is also possible.
However, for $\mu = \mathcal{O}(1-10)$ GeV and $v_\phi \sim 100 \GeV$ which we take in this paper, 
the degeneracy is not high.  In this case, even if $X_R$ might not decay until $X_I$ freezes out, 
the number density of $X_R$ is much smaller than that of $X_I$.
Hence, we can ignore the possible effect of co-annihilation. 
For $\delta m \equiv m_R - m_I \gg m_{Z'}$, the decay rate of $X_R$ is 
\beq \label{GammaR}
\Gamma_R 
%&=& \frac{\alpha_X}{4} \l( \frac{m_R}{m_{Z'}} \r)^2 m_R
%\nonumber \\
%&\times& \l[ 1 - \frac{\l( m_I + m_{Z'} \r)^2}{m_R^2} \r]^{3/2} \l[ 1 - \frac{\l( m_I - m_{Z'} \r)^2}{m_R^2} \r]^{3/2}
%\nonumber \\
%&\approx& 
\approx \frac{\alpha_X}{4} \l( \frac{m_R}{m_{Z'}} \r)^2 m_R \l[ 1 - \frac{m_I^2}{m_R^2} \r]^3
%\nonumber \\
%&=& 
= \frac{\sqrt{2}}{2} \frac{\mu^2 v_\phi}{m_R^2}
\eeq
Hence, unless $\mu$ is smaller than $\GeV$ scale by many orders of magnitude, $X_R$ decays well before its would-be freeze-out.
Note that, if the mass splitting between $X_R$ and $X_I$ were given by hand, $\Gamma_R$ would diverge in the limit of $m_{Z'}=0$ (or $v_\phi=0$), but in our local $Z_2$ model such 
a divergence is absent.

\subsection{Indirect detection: $\GeV$ scale $\gamma$-ray excess at Fermi-LAT}
%\paragraph{$\GeV$ scale $\gamma$-ray excess at Fermi-LAT and $511 \keV$ line at INTEGRAL/SPI}
In Ref.~\cite{Ko:2014gha}, some of present authors showed that DM pair-annihilations to light non-SM Higgses ($\phi$) which eventually decay dominantly to $b\bar{b}$ or $\tau\bar{\tau}$ can explain the $\GeV$ scale $\gamma$-ray excess in the direction of the Galactic Center (GC) if $\langle \sigma v \rangle_{\phi\phi} \sim 10^{-26} {\rm cm}^3/{\rm s}$ \cite{Goodenough:2009gk,Hooper:2010mq,Hooper:2011ti,Abazajian:2012pn,Hooper:2013rwa,Gordon:2013vta,Huang:2013pda,Abazajian:2014fta,Daylan:2014rsa} (see also \cite{Alvares:2012qv,Okada:2013bna,Alves:2014yha,Berlin:2014tja,Agrawal:2014una,Izaguirre:2014vva,Ipek:2014gua,Abdullah:2014lla,Basak:2014sza,Berlin:2014pya,Cline:2014dwa,Wang:2014elb,Ko:2014loa}).
The model at hand in this paper can work in the same way for the $\gamma$-ray excess as long
as we take \footnote{ 
Recently one of the present author showed that the best fit to the $\gamma$-ray excess 
is achieved when $m_{X_I} \simeq 95.0$GeV and $m_{H_2} \simeq 86.7$GeV with 
$\langle \sigma v \rangle_{\phi\phi} \sim  4 \times10^{-26} {\rm cm}^3/{\rm s}$ and the corresponding 
p-value equal to $\simeq 0.40$ ~\cite{Ko:2015ioa}. The present model can accommodate
such value without any difficulty, although we do not elaborate on this detail.}
\beq \label{mphi-range}
\frac{m_h}{2} < m_I \lesssim 80 \GeV, \ \frac{m_I - m_\phi}{m_I} \ll \mathcal{O}(0.1) .
\eeq
Alternatively, DM annihilation to $Z'$s ($X_I X_I \to Z' Z'$) with $m_{Z'}$ replacing $m_\phi$ in \eq{mphi-range} can do the similar job \cite{Berlin:2014pya,Cline:2014dwa}.

As discussed in Ref.~\cite{Ko:2014gha}, contrary to singlet fermion DM, our scalar dark matter allows $s$-wave annihilations mediated by scalar particles.
This means that in our scenario DM annihilation directly to SM particles might be another possibility to explain the $\gamma$-ray excess from GC too for $30 \GeV \lesssim m_X \lesssim 40 \GeV$.
However we found that the relevant parameter space does not satisfy the bound from the direct detection of dark matter that is discussed in the next section.

%The DM annihilation cross section responsible for the Galactic $\GeV$ scale $\gamma$-ray excess might have to be a little bit smaller than the canonical value for thermal relic, and additional annihilation might have to be introduced. 
%In our scenario, this is achieved naturally due to the existence of dark photon.

\subsection{Direct detection}
In the local $Z_2$ model presented in this letter, the direct detection cross section for the DM does not apply for the dark photon $t-$channel exchange, since it is always inelastic ($X_I N \rightarrow X_R N$) and does not take place for $\delta m \gg E_{\rm kin}$. 
Also, the elastic scattering via virtual excited state is totally negligible for the parameter set of our interest \cite{Batell:2009vb}.
Therefore, the kinetic mixing $\epsilon$ is not constrained by direct detection experiments,  in sharp contrast 
with the unbroken $U(1)_X$ case which was studied in Ref.~\cite{Baek:2013qwa} in great detail. 

In addition, even if Higgs exchange of DM-nucleon scattering is potentially crucial to 
constrain our local $Z_2$ scalar DM model, the existence of extra scalar boson mediating 
dark and visible sectors via Higgs portal interaction(s) has a significant effect on direct 
searches if the mass of the extra non-SM Higgs is not very different from that of SM Higgs  
~\cite{Baek:2011aa,Baek:2012se}, and the constraint from direct searches can be 
satisfied rather easily.  Note that this feature is not captured at all in the global $Z_2$ 
scalar DM model where dark Higgs (and also dark photon, although it is irrelevant here) 
is absent  \footnote{See Refs.~\cite{Baek:2011aa,Baek:2012se} for the original discussions 
on this point, and Ref.~\cite{HiggsPortalEFT}  for more discussion on the correlation between 
the invisible Higgs branching ratio and the direct detection cross section in the Higgs portal 
SFDM and SVDM models.}.

The Higgs mediated spin-independent elastic 
%scattering, the differential cross section of the spin-independent 
DM-nucleon scattering is given by 
%\bea
%\frac{d \sigma_p^{\rm SI}}{d \Omega} 
%&=& \frac{m_{\rm r}^2}{\pi^2} \l( \frac{m_p}{4 m_X v_H} \r)^2
%\nonumber \\
%&& \times \l( - \frac{s_\alpha \lambda_1}{q^2 - m_1^2} + \frac{c_\alpha \lambda_2}{q^2 - m_2^2} %\r)^2 f_p^2
%\eea
\bea
\sigma_p^{\rm SI} 
&=& \frac{m_{\rm r}^2}{4\pi} \l( \frac{m_p}{m_X} \r)^2 \frac{c_\alpha^4}{m_1^4}  f_p^2
\\
&\times& \l[ \lambda_{\rm eff} \frac{v_\phi}{v_H} t_\alpha \l( 1 - \frac{m_1^2}{m_2^2} \r) - \lambda_{HX} \l( t_\alpha^2 + \frac{m_1^2}{m_2^2} \r) \r]^2
\nonumber 
\eea
where %the subscript $i$ represents quantity associated with either $H_1$ or $H_2$, 
$m_{\rm r} = m_X m_p / \l( m_X + m_p \r)$, 
%\bea \label{lambda1}
%\lambda_1 &=& \l( \lambda_{\phi X} v_\phi - \sqrt{2} \mu \r) c_\alpha - \lambda_{HX} v_H s_\alpha
%\\ \label{lambda2}
%\lambda_2 &=& \l( \lambda_{\phi X} v_\phi - \sqrt{2} \mu \r) s_\alpha + \lambda_{HX} v_H c_\alpha
%\eea
$f_p \simeq 0.326$ \cite{Young:2009zb} (see also Ref. \cite{Crivellin:2013ipa} for more recent analysis), 
%Ignoring momentum transfer, 
% and assuming the $\lambda_{HX}$-terms in \eqs{lambda1}{lambda2} are sub-dominant, 
%one finds
%\bea
%\sigma_p^{\rm SI} 
%&=& \frac{m_{\rm r}^2}{4\pi} \l( \frac{m_p}{m_X} \r)^2 \frac{c_\alpha^4}{m_1^4}  f_p^2
%\nonumber \\
%&\times& \l[ \lambda_{\rm eff} \frac{v_\phi}{v_H} t_\alpha \l( 1 - \frac{m_1^2}{m_2^2} \r) - %\lambda_{HX} \l( t_\alpha^2 + \frac{m_1^2}{m_2^2} \r) \r]^2
%\eea
%where
and $\lambda_{\rm eff} \equiv ( \lambda_{\phi X} - \sqrt{2} \mu/v_\phi )$.
Currently, the most stringent constraint is from LUX~ \cite{Akerib:2013tjd}, and we may take the bound as $\sigma_p^{\rm SI} < 7.6 \times 10^{-46} {\rm cm}^2$ for $30 \GeV \lesssim m_I, m_\phi \lesssim 80 \GeV$.

\begin{figure}[htp]
\includegraphics[width=0.40\textwidth]{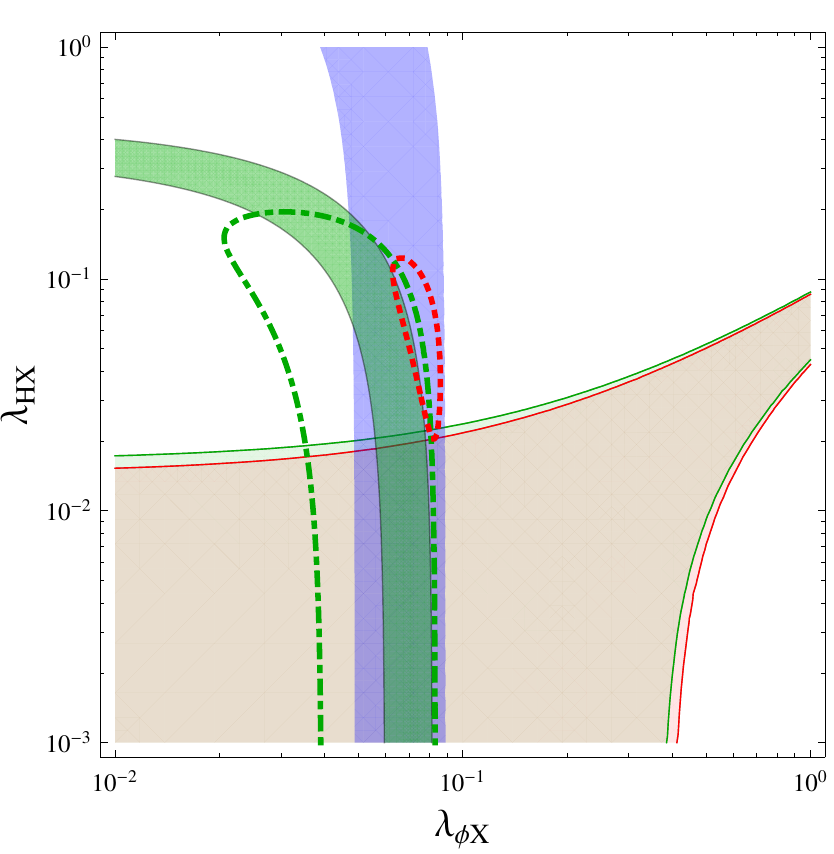}
\caption{\label{fig:sp}
Parameter space for $m_I=80, \ m_\phi = 75 \GeV$ with $\alpha=0.1, \ v_\phi = 100 \GeV$, satisfying constraints from LUX direct search experiment (Green region between thin green lines: $\mu=5\GeV$. Red region between thin red lines: $\mu=7 \GeV$), $\langle \sigma v_{\rm rel} \rangle_{\rm tot} / \langle \sigma v_{\rm rel} \rangle_{\rm th} = 1$ (Dot-dashed green line: $\mu=5\GeV$. Dotted red line: $\mu=7\GeV$), and $1/3 \leq \langle \sigma v_{\rm rel} \rangle_{\phi\phi} / \langle \sigma v_{\rm rel} \rangle_{\rm th} \leq 1$ (Blue region).
In the dark green region, $\langle \sigma v_{\rm rel} \rangle_{Z'Z'} / \langle \sigma v_{\rm rel} \rangle_{\rm th} \leq 0.1$, so the contribution of $Z'$-decay to $\GeV$ scale excess of $\gamma$-ray may be safely ignored.
}
\end{figure}
%-----------------------
In Fig.~3, we show parameter space satisfying the direct detection constraint from LUX, and providing a right amount of relic density for $m_I =80 \GeV$ and $m_\phi=75\GeV$ as an example with a couple of choices of $\mu$.
Also, depicted is the region in which $\GeV$ scale excess of $\gamma$-ray from the GC can 
be explained by $X_I X_I \to \phi \phi$ while $X_I X_I \to Z'Z'$ is somewhat suppressed.
Note that, depending on $\mu$, parameters satisfying $\langle \sigma v_{\rm rel} \rangle_{\rm tot} / \langle \sigma v_{\rm rel} \rangle_{\rm th} = 1$ define a contour in the $(\lambda_{\phi X}, \lambda_{HX})$ plane.
The reason of this is clear from Fig.~\ref{fig:svgg} in which upper bounds of $\lambda_{\phi X}$ and 
$\lambda_{HX}$ can be found.
From \eqss{sv-darkphoton}{lambda1}{lambda2}, one can see that the parameter $\lambda_{\phi X}$ is 
bounded from both above and below  when $\lambda_{HX}$ is very small.
As $\mu$ becomes large, the bounds of $\lambda_{\phi X}$ move toward larger values, and then 
$\lambda_{HX}$ is bounded from below (red dotted line in Fig.~\ref{fig:sp}) because of 
$\langle \sigma v_{\rm rel} \rangle_{\phi \phi}$ contribution.
We found that a region in which all the constraints are satisfied and $\gamma$-ray excess can be explained 
appears for $\mu \sim 5 \GeV$ with $\lambda_{\phi X} \lesssim 0.1$ and $\lambda_{HX} \lesssim 0.01$. 
Although we haven't shown explicitly in this letter, for $m_I \sim 30 \GeV$, we could find a 
parameter space satisfying LUX bound, but $\GeV$ excess of $\gamma$-ray could not be explained 
due to the smallness of $\langle \sigma v_{\rm rel} \rangle_{f\bar{f}}$ contribution to 
$\langle \sigma v_{\rm rel} \rangle_{\rm tot}$.

\section{Implications on collider experiments}
For the canonical set of parameters used in Figs.~\ref{fig:svgg} and~\ref{fig:sp}, SM-Higgs can not decay 
directly  either to dark matter or to dark Higgses.
However, as discussed in Ref.~\cite{Baek:2011aa}, the presence of dark Higgs boson which mixes with the 
SM Higgs boson causes a universal suppression of the signals of SM-channels in collider experiments.
Also, because the mass of dark Higgs is not very different from that of SM Higgs, the mono-jet search is also affected (see Ref.~\cite{pko}), compared with the Higgs-portal models in effective field theory approach.
Since these effects are generic in models of dark Higgs mixed with SM Higgs, it is difficult to probe our 
model at collider even if afore-mentioned effects are found.

\section{Conclusion}

In this letter, we presented a scalar DM model where a local $Z_2$ symmetry originating 
from a spontaneously broken local $U(1)_X$ guarantees the DM stability.
Contrary to the usual global $Z_2$ scalar DM model, our model contains three new extra fields 
(dark photon $Z_\mu^{'}$, dark Higgs $\phi$ and the excited DM partner $X_R$) with kinetic 
and Higgs portal interactions dictated by local gauge invariance and renormalizability. 
Analyzing this model, we showed that the existence of those three extra fields results in 
dark matter phenomenology which is qualitatively different from the usual $Z_2$ scalar DM 
models.  The resulting new model can accommodate thermal relic density of scalar DM without 
conflict with the invisible Higgs branching ratio and the bounds from DM direct detections,  thanks to 
the newly opened channels $X_I X_I \rightarrow Z^{'} Z^{'}, \phi\phi$.   
In particular, the dark Higgs boson allows for the model to accommodate the $\GeV$ scale excess 
of $\gamma$-rays from the direction of Galactic Center. 

We considered the GC $\gamma$-ray for phenomenological analysis of the local $Z_2$ scalar DM model, which depended only 
on a particular corner of parameter space of the model. 
Even if some of these anomalies go away, the local $Z_2$ model presented here could be 
regarded as an alternative to the usual real scalar DM model defined by Eq. (1) 
with global $Z_2$ symmetry. 
The local $Z_2$ model has many virtues: 
(i) dynamical mechanism for stabilizing scalar DM is there with massive dark photon and opens new channels for DM annihilation, 
(ii) DM self-interaction could be accommodated due to the new fields in the local $Z_2$ model~\cite{Ko:2014nha}, 
(iii) the dark Higgs improves EW vacuum stability up to Planck scale~\cite{Baek:2011aa,Baek:2012uj,Baek:2012se}, 
and opens a new window for Higgs inflation~\cite{Ko:2014eia}, 
(iv) the excited DM $X_R$ is built in the model due to $U(1)_X \rightarrow Z_2$ dark symmetry breaking. 
All of these facts make the local $Z_2$ model interesting and DM phenomenology becomes
very rich due to the underlying local dark gauge symmetry stabilizing the scalar DM.  
We plan to present more extensive phenomenological analysis of local $Z_2$ scalar DM model
in separate publications, along with phenomenology of the excited DM and also the local 
$Z_2$ fermion DM model.

\section{Acknowledgement}
This work is supported in part by National Research Foundation of Korea (NRF) Research 
Grant 2012R1A2A1A01006053 (SB, PK, WIP), and by the NRF grant funded by the Korea 
government (MSIP) (No. 2009-0083526) through  Korea Neutrino Research Center 
at Seoul National University (PK).

\end{document}